\documentclass[12pt, a4paper]{article}

\pdfoutput=1

\usepackage{amsmath}
\usepackage{amsfonts}
\usepackage{amssymb}
\usepackage{bbm}
\usepackage{verbatim}
\usepackage{dsfont}
\usepackage{booktabs}
\usepackage{slashed}

\usepackage{epsfig}
\usepackage{color}
\usepackage[table]{xcolor}
\usepackage{graphicx}
\usepackage[]{caption}
\usepackage[listofformat=empty,subrefformat=empty]{subfig}

\usepackage{float}
\usepackage{overpic}

\usepackage{enumerate}
\usepackage{hhline}
\usepackage{multirow}

\usepackage{cite}
\usepackage{xspace}
\usepackage{setspace}

\usepackage[pdftitle={de Sitter vacua and supersymmetry breaking in six-dimensional flux compactification},
  pdfauthor={Wilfried Buchmuller, Markus Dierigl, Fabian Ruehle, Julian Schweizer},
  pdfsubject={orbifolds, effective action, flux, supergravity},
  bookmarksopen, bookmarksnumbered, bookmarksopenlevel=2, colorlinks=false, linkcolor=blue, citecolor=blue, urlcolor=blue]{hyperref}

\newcommand{\Mp}{M_{\mathrm{P}}}
\renewcommand{\tfrac}{\genfrac{}{}{}1}

\newcommand{\Wb}{\ensuremath{\overline{W}}}
\newcommand{\jb}{\ensuremath{\bar{\jmath}}}

\newcommand{\gravb}{\ensuremath{\overline{\psi}}}
\newcommand{\sigb}{\ensuremath{\overline{\sigma}}}
\newcommand{\Xb}{\ensuremath{\overline{X}}}

\begin{document}

\thispagestyle{empty}

\begin{flushright}
DESY-16-022\\
\end{flushright}
\vskip .8 cm
\begin{center}
{\Large {\bf de Sitter vacua and supersymmetry breaking
\\[8pt] in six-dimensional flux compactifications
}}\\[10pt]

\bigskip
\bigskip 
{
{\bf{Wilfried Buchmuller}\footnote{E-mail: wilfried.buchmueller@desy.de}},
{\bf{Markus Dierigl}\footnote{E-mail: markus.dierigl@desy.de}},  
{\bf{Fabian Ruehle}\footnote{E-mail: fabian.ruehle@desy.de}},
{\bf{Julian~Schweizer}\footnote{E-mail: julian.schweizer@desy.de}}
\bigskip}\\[0pt]
\vspace{0.23cm}
{\it Deutsches Elektronen-Synchrotron DESY, 22607 Hamburg, Germany }\\[20pt] 
\bigskip
\end{center}

\begin{abstract}
\noindent
We consider six-dimensional supergravity with Abelian bulk flux
compactified on an orbifold. The effective low-energy action can be
expressed in terms of $\mathcal{N}=1$ chiral moduli superfields with a gauged
shift symmetry. The $D$-term potential contains two Fayet-Iliopoulos
terms which are induced by the flux and by the Green-Schwarz term
canceling the gauge anomalies, respectively. The Green-Schwarz term
also leads to a correction of the gauge kinetic function which turns
out to be crucial for the existence of Minkowski and de Sitter vacua.
Moduli stabilization is achieved by the interplay of the $D$-term
and a nonperturbative superpotential. Varying the gauge coupling and
the superpotential parameters, the scale of the extra dimensions 
can range from the GUT scale down to the TeV scale.
Supersymmetry is broken by $F$- and $D$-terms, and the scale of gravitino,
moduli, and modulini masses is determined by the size of the 
compact dimensions.
\end{abstract}

\newpage 
\setcounter{page}{2}
\setcounter{footnote}{0}
\tableofcontents

{\renewcommand{\baselinestretch}{1.3}

\section{Introduction}
\label{sec:Introduction} 
The ultraviolet completion of the Standard Model remains a challenging
question. There are strong theoretical arguments for supersymmetry at
high scales and, in connection with gravity and string theory, also
for compact extra dimensions.
But in the absence of any hint for
supersymmetry from the LHC the scale of supersymmetry breaking is completely
unknown, except for a lower bound of $\mathcal{O}(1)$ TeV.

In this connection higher-dimensional theories are of interest which
relate the scale of supersymmetry breaking to the size of compact
dimensions via quantized magnetic flux \cite{Bachas:1995ik}. In the context of the
heterotic string it has been argued that five or six dimensions are a
plausible intermediate step on the way from 10d string theory to a 4d
supersymmetric extension of the Standard Model
\cite{Kobayashi:2004ud,Forste:2004ie,Hebecker:2004ce,Buchmuller:2007qf}, and compactifications to six dimensions (6d) are also very interesting from the
perspective of type IIB string theory and F-theory
\cite{Blumenhagen:2002wn,Taylor:2011wt,Ludeling:2014oba}. Furthermore, 6d theories are interesting from a phenomenological point of view. They can naturally explain the multiplicity of quark-lepton
generations as a topological quantum number of vacua with magnetic flux
\cite{Witten:1984dg} and, when compactified on orbifolds, they provide an appealing
explanation of the doublet-triplet splitting in unified theories
\cite{Kawamura:2000ev,Hall:2001pg,Hebecker:2001wq,Asaka:2001eh}. Orbifold
compactifications with flux combine both virtues \cite{Buchmuller:2015jna},
leading to 4d theories reminiscent of ``split''
\cite{ArkaniHamed:2004fb,Giudice:2004tc} or 
``spread'' \cite{Hall:2011jd} supersymmetry.

Many aspects of 6d supergravity theories have already been studied 
in detail in the past. This includes the complete Lagrangian with
matter and gauge fields \cite{Nishino:1984gk,Nishino:1986dc}, compactification of gauged
supergravity with a monopole background \cite{Aghababaie:2002be}, 
localized Fayet-Iliopoulos terms generated by quantum corrections
\cite{Lee:2003mc}, singular gauge fluxes at the fixed points \cite{vonGersdorff:2006nt}
and the cancellation of
bulk and fixed point anomalies by the Green-Schwarz mechanism
\cite{Erler:1993zy,Asaka:2002my,Scrucca:2004jn,Park:2011wv}.
In particular, it has been shown in \cite{Braun:2006se} how magnetic flux
together with a nonperturbative superpotential can stabilize both
dilaton and volume modulus.

The present paper extends our previous work \cite{Buchmuller:2015eya} where we showed
that  the Green-Schwarz mechanism also cancels the anomalies due to
the chiral zero modes induced by the magnetic flux. We now demonstrate that
the low-energy effective Lagrangian takes the form of an $\mathcal{N}=1$ supergravity 
model for the moduli superfields with a gauged shift symmetry. The
corresponding Killing vectors are induced by the magnetic flux and
also by the Green-Schwarz term, respectively. Furthermore, $\mathcal{N}=1$
supersymmetry implies an important modification of the gauge kinetic
function. As discussed already in \cite{Buchmuller:2016dai}, this allows for a new class
of metastable de Sitter solutions.

The paper is organized as follows. In Section \ref{sec:6dSugra} we derive the K\"ahler
potential, the gauge kinetic function and the $D$-term potential of the
4d theory with special emphasis on the effects of the Green-Schwarz term.
Minkowski and de Sitter vacua are analyzed in Section \ref{sec:Bosons} and it is
shown how dilaton, volume and shape moduli can be stabilized by the flux
together with a nonperturbative superpotential. The $U(1)$ vector
boson mass, the charged scalar mass, the moduli masses, and the axion masses are evaluated for two examples of Minkowski vacua with different size of
the compact dimensions. An important aspect of the model is the
realization of the super-Higgs mechanism with combined $F$- and $D$-term
breaking of supersymmetry. This, together with the modulini masses, is discussed in Section \ref{sec:Fermions}. 
Details of the search for de Sitter vacua and the super-Higgs
mechanism are given in Appendix \ref{parameters} and Appendix \ref{Appfermions}, respectively. 
Appendix \ref{appnumerical} contains more details about the example models, including numerical values for their mass spectra.

\section{Effective supergravity action}
\label{sec:6dSugra}

Let us first consider the bosonic part of the six-dimensional supergravity action with a $U(1)$ gauge
field\footnote{We use the differential geometry conventions of \cite{blumenhagenBasic}, the volume form multiplying $R$ is understood.},
\begin{align}
S_B =\int\left(\frac{M_6^4}{2} \left( R - d \phi \wedge \ast d \phi
  \right) - \frac{1}{4 M_6^4 g_6^4} e^{2 \phi} H \wedge \ast H -
  \frac{1}{2g_6^2} e^\phi F \wedge \ast F\right)\,,\label{Sb}
\end{align}
involving the Ricci scalar $R$, the dilaton $\phi$, the gauge field $A
= A_M \, dx^M$ and the antisymmetric tensor field $B = \tfrac{1}{2}
B_{MN} \, dx^M \wedge dx^N$. The corresponding fields strengths are
given by
\begin{align}
F = dA \,, \quad H = dB - X^0_3 \,;
\end{align}
$M_6$ is the 6d Planck mass and $g_6$ denotes the 6d gauge coupling of mass dimension $-1$. 
The 3-form $X^0_3$ is the difference between the Chern-Simons forms
$\omega_{3L}$  and $\omega_{3G}$ for the spin connection $\omega$ and the
gauge field $A$, respectively. In the following we ignore
$\omega_{3L}$ since we will not discuss gravitational anomalies, i.e.,
\begin{align}
X_3^0  = - \omega_{3G}  = - A\wedge F\,.
\end{align}
We choose as background geometry the product space $M \times T^2/\mathbb{Z}_2$ with the metric
\begin{align}
(g_6)_{MN} = \begin{pmatrix} r^{-2} (g_4)_{\mu \nu} & 0 \\ 0 & r^2
  (g_2)_{mn} \end{pmatrix} \,,
\label{6dmetric}
\end{align}
where $\mu,\nu = 0 \dots 3$ correspond to the 4d Minkowski space and
$m,n = 5,6$ to the internal space.
It is convenient to use dimensionless coordinates for the compact
space, $(x^5,x^6) = (y_1,y_2) L$, where $L$ is a fixed physical
length scale. The rescaling by the dimensionless radion field $r$ in \eqref{6dmetric} leads
to standard kinetic terms for the moduli. 
The shape of the internal space is parametrized
by the two real shape moduli $\tau_{1,2}$ in the two-dimensional metric $(g_2)_{mn}$,
\begin{align}
(g_2)_{mn} = \frac{1}{\tau_2} \begin{pmatrix} 1 & \tau_1 \\ \tau_1 & \tau_1^2 + \tau_2^2 \end{pmatrix} \,,
\end{align}
and the orbifold projection acts as $x_m \rightarrow -x_m$. The
physical volume of the internal space is $V_2 = \tfrac{1}{2} \langle r
\rangle^2 L^2$, where $\langle r \rangle$ is the vacuum expectation
value of the radion field $r$.

Neglecting the gravitational backreaction on the geometry of the internal space, a constant bulk flux is a
solution of the equations of motion,
\begin{align}
\langle F \rangle = \frac{f}{L^2} v_2\,, \quad f = \text{const} \,, 
\end{align}
where $v_2 = dy_1\wedge dy_2$.
Furthermore, we add to the gauge-gravity sector a bulk hypermultiplet
containing a 6d Weyl fermion with charge $q$ and two complex
scalars. The hypermultiplet can be decomposed into two 4d $\mathcal{N}=1$ chiral
multiplets with charges $q$ and $-q$, respectively. 
The two complex scalars, $\phi_+$ and $\phi_-$, have gauge interactions
and a scalar potential which, in 4d $\mathcal{N}=1$ language, corresponds to an
$F$-term and a $D$-term potential of the two chiral multiplets \cite{ArkaniHamed:2001tb}, 
\begin{equation}
\begin{split}
S_M = -\int \Big(&(d + iqA)\phi_+\wedge \ast (d - iqA)\bar{\phi}_+ + (d -
iqA)\phi_-)\wedge \ast (d + iqA)\bar{\phi}_-) \\
& + 2g_6^2 q^2e^{-\phi} |\phi_+\phi_-|^2 + \frac{g_6^2q^2}{2}e^{-\phi}(|\phi_+|^2 - |\phi_-|^2)^2\Big)\,.
\end{split}
\end{equation}
Due to charge quantization the value of the background flux can only take discrete values,
\begin{align}
\frac{q}{2\pi} \int_{T^2/\mathbb{Z}_2} \langle F \rangle = \frac{qf}{4\pi} \equiv -N \in \mathbb{Z} \,.
\label{orbquantization}
\end{align}
For $|N|>0$, the index theorem guarantees the presence of $N$ massless
left-handed 4d Weyl fermions
\cite{Witten:1984dg}.
This model is anomalous, with bulk and fixed point anomalies
calculated in \cite{Asaka:2002my},
\begin{align}\label{6Danomaly}
\mathcal{A} = \Lambda F \wedge \left(\frac{\beta}{2} F\wedge F +
  \alpha \delta_O F\wedge v_2 \right)\,,
\end{align}
where $\beta = -q^4/(2\pi)^3$, $\alpha = q^3/(2\pi)^2$, and
\begin{align}
\delta_O(y) = \frac{1}{4}\sum_{i=1}^4 \delta (y-\zeta_i) \,,
\end{align}
where the $\zeta_i$ correspond to the four fixed points on the orbifold. From the first term in Eq.~\eqref{6Danomaly} it is obvious that the background flux contributes to the chiral anomaly. As shown in \cite{Buchmuller:2015eya}, all these anomalies are canceled
by the Green-Schwarz term
\begin{align}\label{GS1}
S_\text{GS} = -\int \left(\frac{\beta}{2} A\wedge F + \alpha
  \delta_O A\wedge v_2\right) \wedge dB\,.
\end{align}

It is now straightforward to compute the 4d effective action by means of dimensional reduction. Matching the
Ricci scalars and the gauge kinetic terms yields for the tree level 4d Planck mass and the 4d gauge coupling, respectively,
\begin{align}
M_4^2 = \frac{L^2}{2} M_6^4\,, \quad \frac{1}{g_4^2} = \frac{L^2}{2g_6^2} \,.
\end{align}

The gauge part of the 4d effective action has been worked out in \cite{Buchmuller:2015eya}. The field strength $H$ of the antisymmetric
tensor $B$ can be written as
\begin{align}
H = \left(g_4^2 M_4^2 db + 2f \hat{A}\right) v_2 + \hat{H}\,, \quad
\hat{H} = d\hat{B} + \hat{A} \wedge \hat{F} \,,
\end{align}
where $b$, $\hat{A}$ and $\hat{B}$ denote 4d scalar, vector and tensor fields. Trading $\hat{H}$ for the
dual scalar $c$ by means of the Lagrange multiplier term
\begin{align}
\Delta S_\text{cH} = \frac{1}{2g_4^2}\int_M c\ d(\hat{H} - \hat{A}\wedge \hat{F})\,, 
\end{align}
replacing radion and dilaton by the moduli fields $s$ and $t$,
\begin{align}
t = r^2 e^{-\phi}\,, \quad s = r^2 e^\phi\,, 
\end{align}
rescaling the lowest state of the matter field, $\phi_+ \rightarrow \sqrt{2}/L \, \phi_+$,
and dropping the `hat' for the 4d vector field, one obtains
\begin{equation}
\begin{split}
\label{LB}
S^{(4)}_\text{B} = \int_M & \left\{ \frac{M_4^2}{2}\Big(R_4 -\frac{1}{2t^2} dt \wedge \ast dt  
-\frac{1}{2s^2} ds\wedge \ast ds  -\frac{1}{2\tau_2^2} d\tau \wedge \ast d\overline{\tau}\Big) \right.\\  
&-\frac{1}{2g_4^2}\Big( sF\wedge\ast F + (c +g_4^2\beta\bar{\ell}^2b)F\wedge F\Big)
- \frac{g^2 M_4^4}{2st^2} \frac{f^2}{\bar{\ell}^4} \\
&-\frac{M_4^2}{4t^2} \Big(db + \frac{2f}{\bar{\ell}^2} A\Big)\wedge\ast
\Big(db + \frac{2f}{\bar{\ell}^2} A\Big)\\
&-\frac{M_4^2}{4s^2} \Big(dc +g_4^2(2\alpha +\beta f)A )\wedge \ast
(dc + g_4^2(2\alpha +\beta f)A\Big)\\
& \left. -(d + iqA)\phi_+\wedge \ast (d - iqA)\bar{\phi}_+ - \tilde{m}_+^2|\phi_+|^2
- \frac{g_4^2q^2}{2s}|\phi_+|^4 \right\}\,.
\end{split}
\end{equation}
For convenience,
we have introduced the dimensionless parameter
\begin{align}
\bar{\ell} = g_4 M_4 L\,,
\end{align}
and we have dropped\footnote{On the orbifold
  without flux $\phi_-$ is projected out. In the case with flux it
  belongs to the first exited Landau level ($n=1$). Its tree-level mass is
  degenerate with the lowest level ($n=0$) of $\phi_+$, but for the
  following discussion $\phi_-$ is irrelevant.} the matter field $\phi_-$.
Note that the 6d Ricci scalar contains the 4d Ricci
scalar $R_4$ and the kinetic terms of the moduli fields $s$, $t$ and
$\tau \equiv \tau_1+i\tau_2$. For $\phi_+$, the scalar corresponding to the 4d
zero mode $\psi_+$, the flux generates the mass term
\begin{align}\label{m2class}
\tilde{m}_+^2 = -g_4^2 M_4^2 \frac{qf}{st\bar{\ell}^2}\,.
\end{align}

For a vacuum expectation value $\langle r \rangle \neq 1$ a
constant Weyl rescaling $(g_4)_{\mu\nu} \rightarrow \langle r \rangle^2
(g_4)_{\mu\nu}$ has to be performed such that the rescaled metric
describes physical 4d distances. The Ricci scalar of the rescaled
metric is then multiplied by $M^2_\text{P} = M^2_4 \langle r \rangle^2$.
$M_\text{P}$ corresponds to the physical Planck mass and is related to
$M_6$ by the physical volume, $M^2_\text{P} = V_2 M_6^4$. Analogously, the
physical coupling $g = g_4/\langle r \rangle$ is related to $g_6$ by
$g^{-2} = V_2 g^{-2}_6$.
Furthermore, we now rescale moduli and matter fields,
$(s,t,\ldots) \rightarrow (s,t,\ldots) \langle r \rangle^2$, $\phi_+ \rightarrow \phi_+/\langle r \rangle$. The
resulting final 4d bosonic action is identical to Eq.~\eqref{LB}
except for a change of parameters,
\begin{equation}
\begin{split}
\label{LBfinal}
S^{(4)}_\text{B} = \int_M & \left\{ \frac{M_\text{P}^2}{2}\Big(R_4 -\frac{1}{2t^2} dt \wedge \ast dt  
-\frac{1}{2s^2} ds\wedge \ast ds  -\frac{1}{2\tau_2^2} d\tau \wedge \ast d\overline{\tau}\Big) \right.\\  
&-\frac{1}{2g^2}\Big( sF\wedge\ast F + (c +g^2\beta\ell^2b)F\wedge F\Big)
- \frac{g^2 M_\text{P}^4}{2st^2} \frac{f^2}{\ell^4} \\
&-\frac{M_\text{P}^2}{4t^2} \Big(db + \frac{2f}{\ell^2} A\Big)\wedge\ast
\Big(db + \frac{2f}{\ell^2} A\Big)\\
&-\frac{M_\text{P}^2}{4s^2} \Big(dc +g^2(2\alpha +\beta f)A )\wedge \ast
(dc + g^2(2\alpha +\beta f)A\Big)\\
& \left. -(d + iqA)\phi_+\wedge \ast (d - iqA)\bar{\phi}_+ - m_+^2|\phi_+|^2
- \frac{g^2q^2}{2s}|\phi_+|^4 \right\}\,,
\end{split}
\end{equation}
where $M_\text{P}$ and $g$ are physical Planck mass and gauge coupling,
respectively. The  parameter $\bar{\ell}$ is replaced by
\begin{align}
\ell = \langle r \rangle \bar{\ell} = g_4 M_4 \langle r \rangle L = g M_\text{P} \langle r \rangle L \,,
\end{align}
and the scalar mass is given by
\begin{align}\label{m2classphysical}
m_+^2 = \langle r \rangle^2 \tilde{m}_+^2 = -g^2 M_\text{P}^2
\frac{qf}{st\ell^2} = \frac{qf}{2st V_2}\,.
\end{align}
By construction, the rescaled moduli fields satisfy $\langle st \rangle = 1$. We
conclude that given a vacuum field configuration $((g_4)_{\mu\nu}, \langle s
\rangle, \langle t \rangle)$ one can always perform a rescaling such
that the new metric $(g_4)_{\mu\nu}$ describes physical 4d distances and the new
moduli fields satisfy $\langle st \rangle = 1$. The length scale
$\langle r \rangle L$ corresponds to the physical size of the extra
dimensions in terms of the physical Planck mass. In the following we
shall therefore directly search for vacua with $\langle r \rangle =1$,
and we set $M_\text{P} =1$.

The flux compactification on the orbifold $T^2/\mathbb{Z}_2$ should lead to
a 4d theory with spontaneously broken $\mathcal{N}=1$ supersymmetry. Indeed, introducing the complex fields
\begin{align}
S = \tfrac{1}{2}(s + ic)\,,\quad T = \tfrac{1}{2}(t + ib)\,,\quad U=
\tfrac{1}{2} (\tau_2 + i\tau_1)\,,
\end{align}
and comparing expression \eqref{LB} with the standard $\mathcal{N}=1$ supergravity
Lagrangian \cite{Wess:1992cp}, one immediately confirms that the
kinetic terms of the moduli and the matter field are reproduced by the
K\"ahler potential\footnote{We use the same symbol for a chiral superfield and the related
  complex scalar; the 4d vectorfield $A$ is contained in the real superfield $V$.}
\begin{align}
K = - \ln(S + \overline{S} + i X^S V) - \ln (T+\overline{T} + iX^T V) - \ln (U+\overline{U}) +
\bar{\phi}_+ e^{2qV} \phi_+ \,,
\label{kahler}
\end{align}
with the Killing vectors
\begin{align}
X^T = -i\frac{f}{\ell^2}\,, \quad
X^S = -ig^2\alpha(N+1)\,, 
\end{align}
where we have used the relation \eqref{orbquantization} from the flux quantization $\alpha+\beta f/2 =
\alpha (N+1)$; 
the gauge interaction of the matter field corresponds to the Killing vector
\begin{align}
X^+ = -iq\phi_+\,.
\end{align}
From the coefficient of $F\wedge F$ one reads off the gauge kinetic
function
\begin{align}\label{GKF}
H = h_S S + h_T T = 2(S + g^2\beta\ell^2 T) \,,
\end{align}
whose real part we denote by $h$. At first glance this appears to be at variance with the coefficient of $F\wedge\ast F$
suggesting $H=2S$. Note, however, that with the Green-Schwarz term we
have included only part of the one-loop corrections to the effective
action. A complete calculation should preserve supersymmetry. Hence,
we expect that there are further contributions such that
the correct gauge kinetic function is indeed given by
Eq.~\eqref{GKF}, which remains to be confirmed by an explicit
calculation. The one-loop corrected gauge kinetic function depends now on
the moduli fields $S$ and $T$. Such a $T$-dependence has previously
been found in heterotic string compactifications as an effect of quantum
corrections \cite{Ibanez:1986xy,Dixon:1990pc}.

Knowing the Killing vectors and the gauge kinetic function, the $D$-term
potential is given by \cite{Wess:1992cp}
\begin{align}
V_D = \frac{g^2}{2h} D^2\,,
\end{align}
where $h = (s+g^2\beta\ell^2 t)$ and 
\begin{align}
D &= iK_T X^T + iK_S X^S + iK_{\phi_+}X^+ \equiv q|\phi_+|^2 + \xi\,,\\
\xi &\equiv \xi_T + \xi_S = - \frac{f}{t\ell^2} -
\frac{g^2\alpha(N+1)}{s}\,.
\end{align}
This is the standard $D$-term potential for a charged complex scalar
with a field-dependent Fayet-Iliopoulos (FI) term $\xi$. The scalar potential in Eq.~\eqref{LBfinal} contains the
classical part of the $D$-term potential which is given by $\xi_T$. It contributes linearly to the tree-level mass of the charged scalar and quadratically to the energy density.
Supersymmetry requires the quantum correction $\xi_S$ in addition,
analogously to the gauge kinetic function. We therefore keep $\xi_S$ in
the $D$-term potential, which again should be verified by direct calculation. 

Fayet-Iliopoulos terms $\xi_i$ which are generated at the orbifold fixed points by quantum
corrections have been discussed in \cite{Lee:2003mc} in the case of
zero bulk flux. These FI terms are $\mathcal{O}(q)$, their sum and
therefore the effective 4d FI term vanishes ($\sum_i \xi_i = 0$), and they are
locally canceled by a dynamically generated flux that modifies the
zero-mode wave functions.
As we have shown, there is a non-vanishing 4d FI term 
 $\xi_S = \mathcal{O}(q^3)$, which follows from the Green-Schwarz term.
Following \cite{Lee:2003mc}, it would be interesting to analyze
 systematically the interplay between the classical bulk flux and the 
local quantum flux at the fixed points, and to study their joint effect on the zero-mode 
wave functions.

\section{Moduli stabilization and boson masses}
\label{sec:Bosons}

The explicit form of the supersymmetric effective action enables us to determine the masses of the bosons in the theory. For the gauge boson $A$ and the charged scalar field $\phi_+$ the results were already given in \cite{Buchmuller:2015eya}. However, the non-trivial effective gauge coupling additionally includes higher order terms\footnote{Furthermore, the normalization of the field strength $H$ differs compared to that in \cite{Buchmuller:2015eya} in order to match the supergravity conventions.} that so far have not been incorporated.

The vector boson mass can be extracted form the bilinear term of the 4d gauge field in Eq.~\eqref{LB} and reads
\begin{equation}
\begin{split}
m_A^2 &= \frac{2 g^2}{h} \left( \frac{f^2}{\ell^4 t^2} + \frac{g^4\alpha^2 (N+1)^2}{s} \right) \\
&= \frac{1}{2 h} \left( \frac{16 \pi^2 N^2}{(g q)^2 \, V_2^2 \, t^2} + \frac{4 g^6 \alpha^2 (N+1)^2}{s^2} \right)\,,
\label{vectormass}
\end{split}
\end{equation}
The lowest charged scalar mass originates from the $D$-term potential and includes corrections due to anomaly cancellation and non-trivial gauge kinetic function. This modifies the classical scalar mass term \eqref{m2class} to
\begin{equation}
\begin{split}
m_+^2 &= \frac{g^2}{h}
\left(-\frac{qf}{t\ell^2} - \frac{g^2q\alpha(N+1)}{s}\right) \\
&= \frac{1}{2h} \left( \frac{4\pi N}{V_2 \, t} - \frac{2 g^4 q \alpha (N+1)}{s} \right) \,.
\label{chargedmass}
\end{split}
\end{equation}
Without flux the negative second term, accounting for the quantum corrections induced by anomaly cancellation, leads to a vacuum expectation value for $\phi_+$ such that the total $D$-term vanishes. For non-vanishing flux, however, $-qf = 4 \pi N> 0$ and the first term tends to stabilize the scalar field $\phi_+$ at zero. Consequently, the charged scalars are stabilized at the origin as long as the first, flux induced, term in Eq.~\eqref{chargedmass} dominates, i.e.\ for
\begin{align}
s > \frac{g^2 \ell^2 q^4}{(2 \pi)^3} \, \frac{N+1}{2N} \, t\,.
\label{posDterm}
\end{align}
Moreover, the real part of the gauge kinetic function, determining the effective $U(1)$ gauge coupling, has to be positive for the theory to be consistent. Because of the negative prefactor of $t$ this leads to a restriction of the physical moduli space parametrized by $s$ and $t$ (see Eq.~\eqref{GKF}),
\begin{align}\label{posh}
s > \frac{g^2 \ell^2 q^4}{(2 \pi)^3} t \,.
\end{align}
Hence, for $N \geq 1$ the charged fields are always stabilized at zero
for $s$ and $t$ in the physical region of the moduli
space\footnote{Note that in Eqs.~\eqref{vectormass}-\eqref{posh} only
  the product $gq$ appears, since $\ell \propto g$. In the following we set $q=1$.}. Furthermore, in this regime the $D$-term contribution to the scalar potential is positive. However, it is obvious that the runaway-type $D$-term potential alone can not stabilize the moduli fields and we have to include a superpotential.

A superpotential for the moduli can arise at the orbifold fixed points
in the six-dimensional supergravity theory. The superpotential in the
4d effective action is the sum of the fixed point contributions.
Consistency requires this superpotential to be gauge invariant\footnote{Unless one gauges an R-symmetry leading to constant FI-terms \cite{Villadoro:2005yq, Antoniadis:2014hfa}.}. In the following discussion we neglect a possible coupling of moduli to charged bulk fields and restrict our attention to a superpotential that only depends on the moduli fields. For that reason we define a gauge invariant combination of the two shifting chiral superfields $S$ and $T$
\begin{align}
Z = \tfrac{1}{2} (z + i \tilde{c}) \equiv -i X^T S + i X^S T \,.
\end{align}
The superpotential is a holomorphic function of $Z$ and the gauge invariant modulus $U$. Inspired by typical superpotentials induced by nonperturbative effects, such as gaugino condensation or instanton corrections, we assume the following functional dependence
\begin{align}
W = W(Z, U) = W_0 + W_1 \, e^{-a Z} + W_2 \, e^{-\tilde{a} U} \,,
\label{superpot}
\end{align}
where, without loss of generality, we choose the parameters $W_0$, $W_1$, and $W_2$ to be real. For an exponential suppression of these nonperturbative effects we further demand that $a$ and $\tilde{a}$ are real and positive. The $F$-term scalar potential reads
\begin{align}
V_F = e^K \left( K^{i \jb} D_i W D_{\jb} \Wb - 3 |W|^2 \right) \,.
\label{Fpot}
\end{align}
Note that the K\"ahler potential \eqref{kahler} is of the no-scale
form, and the contribution $-3 |W|^2$ is therefore
canceled. The F-term potential \eqref{Fpot} also contributes to the
scalar mass term $m^2_+$. However, this contribution is
$\mathcal{O}(W^2)$, which is much smaller than the leading contribution
$\mathcal{O}(D)$, and it can therefore be neglected.

With the charged fields stabilized at zero the $D$-term
\begin{align}
D = i K_T X^T + i K_S X^S = -\frac{i}{t} X^T - \frac{i}{s} X^S \,,
\label{Dterm}
\end{align}
and the linearly independent combination
\begin{align}
E = i K_T X^T - i K_S X^S = -\frac{i}{t} X^T + \frac{i}{s} X^S \,,
\label{Eterm}
\end{align}
can be used to rewrite the scalar potential in the convenient form
\begin{align}
V = \frac{s t}{2 \tau_2} \left( D^2 + E^2 \right) A + \frac{\tau_2}{s t} \tilde{A} - \frac{1}{\tau_2} E B - \frac{1}{s t} \tilde{B} + \frac{g^2}{2 h} D^2 \,,
\label{scalarpot}
\end{align}
where the parameters of the superpotential are encoded in
\begin{align}
A &= |\partial_Z W|^2 \,,						&\tilde{A} &= |\partial_U W|^2 \,, \nonumber\\
B &= (\partial_Z W) \Wb + W (\partial_{\overline{Z}} \Wb)\,, 	 	&\tilde{B} &= (\partial_U W) \Wb + W (\partial_{\overline{U}} \Wb)\,.
\end{align}
For the specific form \eqref{superpot} of the superpotential these quantities are given in App.~\ref{parameters}.
In order to find minima with vanishing or small cosmological constant one has to solve the four equations
\begin{align}
\partial_S V = 0 \,, \quad \partial_T V = 0 \,, \quad \partial_U V = 0 \,, \quad V = \epsilon \geq 0 \,.
\label{constraints}
\end{align}
These are worked out in App.~\ref{parameters}. We use an inverted procedure to obtain the superpotential parameters by solving Eqs.~\eqref{constraints} after fixing the vacuum expectation values of the moduli fields and the energy density. Consequently, for vanishing or small cosmological constant the derived superpotential is fine-tuned to compensate the large positive $D$-term depending on $g$ and $L$. To solve Eqs.~\eqref{constraints} we further have to fix one of the parameters, which we choose to be $\tilde{a}$. However, one of the above combinations, $A$, can be uniquely determined in terms of the moduli values at the minimum. Up to the prefactor $\tau_2$ the form is identical to the two moduli case discussed in \cite{Buchmuller:2016dai},
\begin{align}
A = \frac{g^2 \tau_2}{2 s t \, h^2(\rho^2 -1)} \left( h_T t \rho + h (2- \rho + \rho^2) + \frac{4 h^2 \epsilon}{g^2 E^2} \right)\,,
\end{align}
where $\rho$ is the ratio $D/E$. Therefore, the arguments for the
existence of vacua given in \cite{Buchmuller:2016dai} carry over to
the three moduli case. Importantly, one necessary condition is that
the prefactor $h_T$ of one of the moduli in the gauge kinetic function
is negative. In the above case this constrains the allowed moduli
region to a regime where the $D$-terms are positive definite. The
additional negative contributions from the $F$-terms of the gauge
invariant superpotential allow to find Minkowski or de Sitter vacua
with all moduli stabilized. In this way we can construct models with Minkowski or de Sitter vacua with a size of the internal space that ranges between GUT and TeV scale.

Given a vacuum with $r = 1$ and the cancellation between $F$- and $D$-term contributions to the potential, we can derive the $L$-dependence of the various parameters and masses. For $(s,t)$ parametrized by $(\kappa, \kappa^{-1})$ solutions to Eqs.~\eqref{constraints} can be found for certain values of the parameter $g^2 L/ \kappa$. Keeping this parameter combination fixed we then obtain models with different size of the extra dimensions. Accordingly, the scaling of the effective gauge coupling and $D$-term potential is
\begin{align}
g_{\text{eff}}^2 \propto L^{-1} \,, \quad V_D \propto L^{-3} \,.
\label{Dtermscaling}
\end{align}
This directly implies a scaling of the superpotential parameters
\begin{align}
W_0, W_1, W_2 \propto L^{-3/2} \,, \quad a \propto L \,,
\label{superpotscaling}
\end{align}
and allows to deduce the behavior of the bosonic masses
\begin{align}
m_+^2 \propto L^{-2} \,, \quad m_A^2 \propto L^{-3} \,, \quad m_{\text{moduli}}^2 \propto L^{-3} \,, \quad  m_{\text{axions}}^2 \propto L^{-3}
\label{bosmassscaling}
\end{align}

\subsubsection*{Superpotential and boson masses: two examples}

In order to illustrate our general results and 
to get some intuition for the parameters and mass scales involved, we
work out explicitly two models exhibiting Minkowski vacua with
different size of the internal dimensions. 
As explained above we start with the choice of the gauge coupling $g$,
the size $L$ of the compact dimensions, and $\tilde{a}$ in the superpotential. The parameters of the two models read
\begin{equation}
\begin{split}
\text{Model I:}& \quad g = 0.2 \,, \hspace{1.2cm} L = 200 \,, \quad \tilde{a} = 2 \,, \\
\text{Model II:}& \quad g = 4 \times 10^{-3} \,, \enspace L = 10^6 \,, \hspace{0.47cm} \tilde{a} = 3 \,.
\label{vacua}
\end{split}
\end{equation}
The number of flux quanta in both vacua is set to $N=3$, which ensures the stabilization of the charged scalar fields, see Eq.~\eqref{posDterm}, and already hints at a multiplicity which can be used in grand unified model building \cite{Buchmuller:2015jna}. The complex shape modulus is stabilized at $\tau = 1$, which corresponds to the square torus assumed in \cite{Buchmuller:2016dai}.

To achieve $r = 1$ in the vacuum we parametrize $(s,t)$ by $\left(\kappa, \kappa^{-1} \right)$ as above. Therefore, the mass scale of the internal dimensions (in 4d Planck units) is given by
\begin{align}
\left(V_2\right)^{-1/2} = \frac{\sqrt{2}}{L} \,, \quad (V_2)_\text{I}^{-1/2} \approx 7.1 \times 10^{-3} \,, \quad  (V_2)_\text{II}^{-1/2} \approx 1.4 \times 10^{-6} \,,
\end{align}
which corresponds to the GUT scale and an intermediate mass scale, respectively. The moduli are stabilized at $\kappa_{\text{I}} = 0.6$, $\kappa_{\text{II}} = 1.2$. As discussed above, the combination $g^2 L/\kappa$ remains constant in both models. The minima in the two different models are plotted in the $s$-$t$ and the $s$-$u$ plane in Figures \ref{minimumplotI} and \ref{minimumplotII}.
\begin{figure}[t]
\centering
	\includegraphics[width = .47 \textwidth]{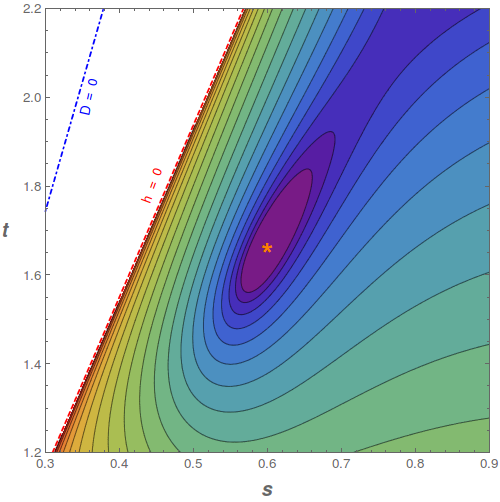}
	\hspace{.04 \textwidth}
	\includegraphics[width = .47 \textwidth]{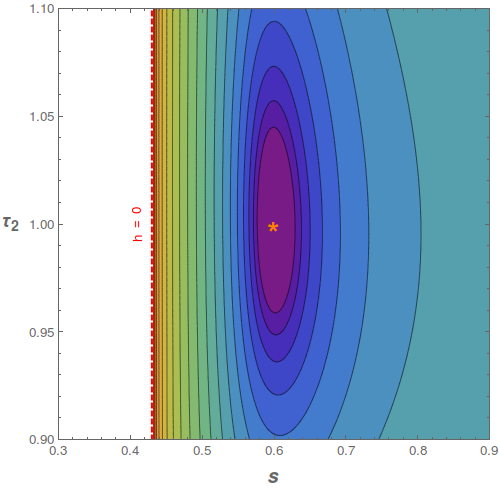}
	\caption{Contour plot of the potential in Model I ($L = 200$) in the $s$-$t$ plane (fixed $\tau_2 = 1$) and $s$-$\tau_2$ plane (fixed $t = \tfrac{5}{3}$).}
	\label{minimumplotI}
\end{figure}
\begin{figure}[t]
\centering
	\includegraphics[width = .47 \textwidth]{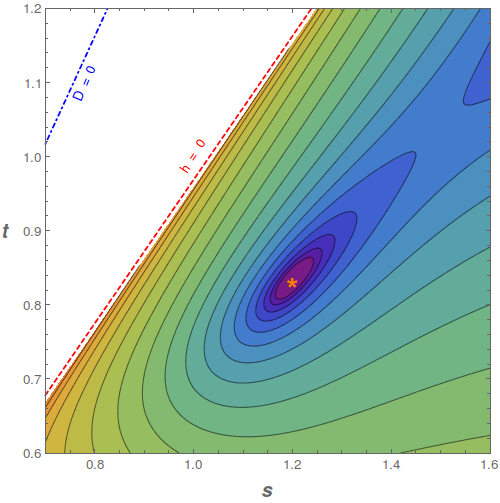}
	\hspace{.04 \textwidth}
	\includegraphics[width = .47 \textwidth]{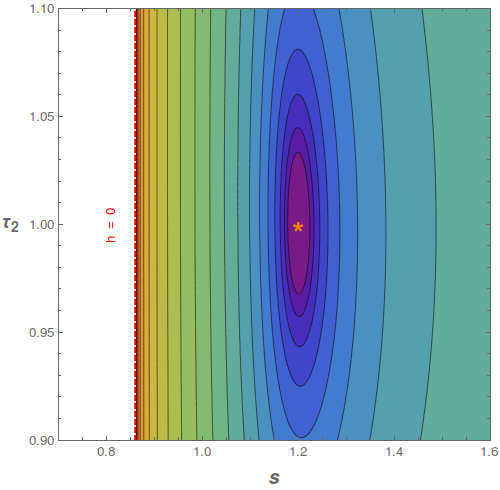}
	\caption{Contour plot of the potential in Model II ($L = 10^6$) in the $s$-$t$ plane (fixed $\tau_2 = 1$) and $s$-$\tau_2$ plane (fixed $t = \tfrac{5}{6}$).}
	\label{minimumplotII}
\end{figure}
One immediate consistency check for the solutions is the value of the effective gauge coupling which has to be positive and small enough to allow a perturbative treatment
\begin{align}
\left(g_{\text{eff}}\right)_{\text{I}} \approx 0.49 \,, \quad \left(g_{\text{eff}}\right)_{\text{II}} \approx 6.9 \times 10^{-3} \,.
\end{align}
This perfectly matches the scaling behavior in Eq.~\eqref{Dtermscaling}. The charged scalar and vector boson masses can then be evaluated numerically using Eqs.~\eqref{vectormass} and \eqref{chargedmass}
\begin{align}
\begin{split}
\left( m_+ \right)_{\text{I}}  & \approx 4.2 \times 10^{-2} \,, \quad \, \left( m_A \right)_{\text{I}} \, \approx  1.1 \times 10^{-2}  \,, \\
\left( m_+ \right)_{\text{II}} & \approx 8.3 \times 10^{-6} \,, \quad \left( m_A \right)_{\text{II}} \approx  3.0 \times 10^{-8} \,.
\end{split}
\end{align}
Again, the scaling with $L$ of Eq.~\eqref{bosmassscaling} is realized. For the masses of the moduli fields $\varphi_i = (s,t,\tau_2)$ and the axions $\tilde{\varphi}_i = (c,b,\tau_1)$ we need to evaluate the superpotential parameters. From Eq.~\eqref{superpotscaling} we would expect a factor of $\mathcal{O}(10^6)$and indeed the respective orders of magnitude are
\begin{align}
(W_0, W_1, W_2)_{\text{I}} \sim \mathcal{O}(10^{-2}) \,, \quad (W_0, W_1, W_2)_{\text{II}} \sim \mathcal{O}(10^{-8}) \,.
\label{superpotpara}
\end{align}
The numerical values are given in App.~\ref{appnumerical}. The nonperturbative exponent in both vacua is $a z \approx 9.4$. Knowing the superpotential, and after canonical normalization, the eigenvalues of the moduli masses matrix can be calculated,
\begin{align}
m_{ij}^2 = \frac{\partial^2 V}{\partial \varphi_i \partial \varphi_j} \Big|_{\langle \varphi_i \rangle, \, \langle \tilde{\varphi}_i \rangle = 0} \,.
\end{align}
The eigenvalues are all of the order of the vector boson mass and slightly larger than the gravitino mass. The scaling between the two models matches the one predicted in Eq.~\eqref{bosmassscaling}. The same is true for the two non-vanishing eigenvalues of the axion mass matrix
\begin{align}
\tilde{m}_{ij}^2 = \frac{\partial^2 V}{\partial \tilde{\varphi}_i \partial \tilde{\varphi}_j} \Big|_{\langle \varphi_i \rangle, \, \langle \tilde{\varphi}_i \rangle = 0} \,,
\end{align}
so that
\begin{align}
m_{\text{moduli}} \sim m_{\text{axions}} \sim m_A > m_{3/2} \,.
\end{align}
One massless axion gives mass to the vector boson via the St\"uckelberg mechanism. Their numerical values in both vacua are discussed in App.~\ref{appnumerical}, where the scaling is explicitly demonstrated.
\begin{figure}[ht!]
	\centering
		\includegraphics[height = .4 \textheight]{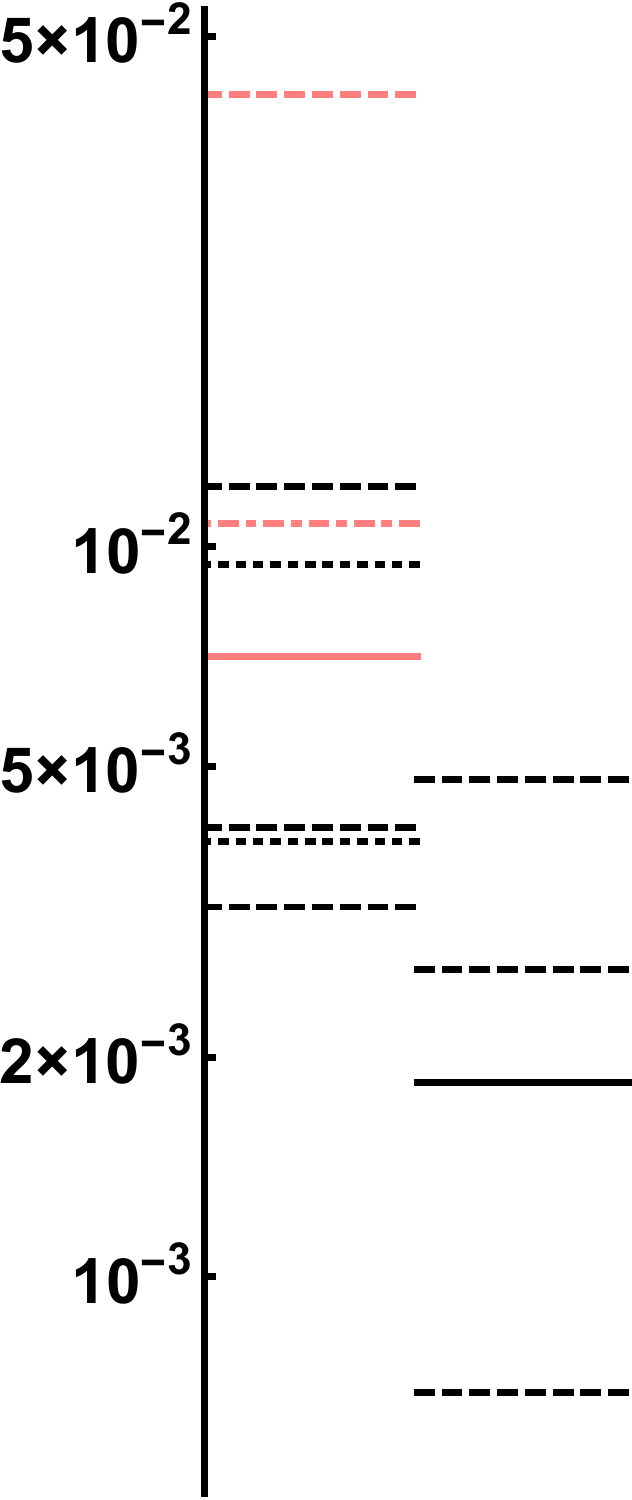}\hspace{1cm}
		\includegraphics[height = .3 \textheight]{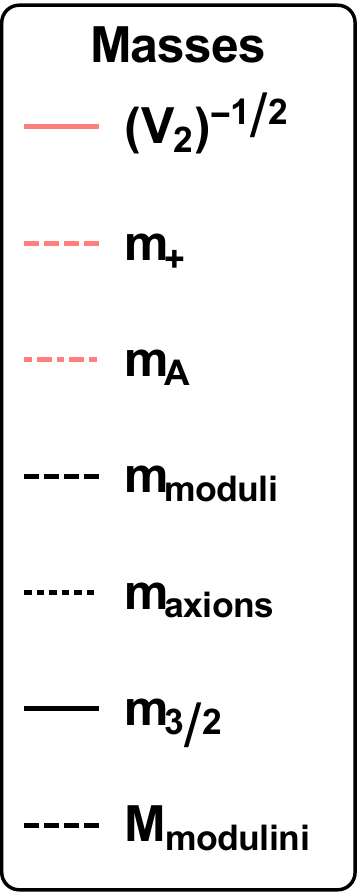}
\hspace{1cm}
		\includegraphics[height = .4 \textheight]{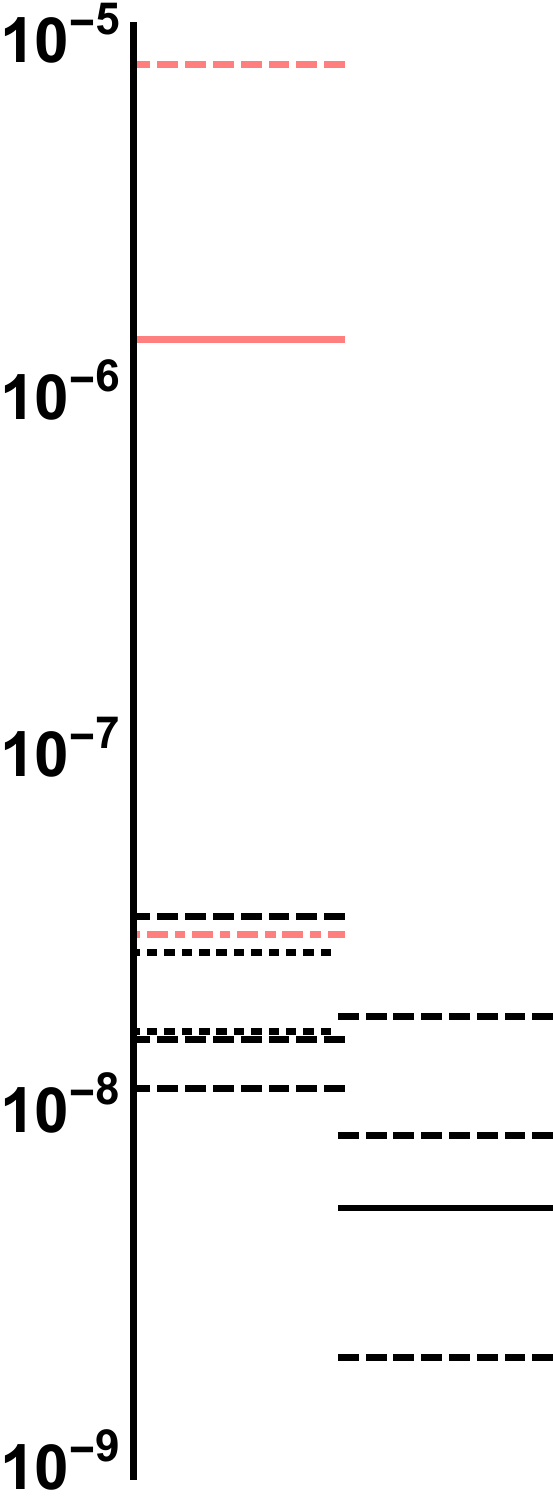}
	\caption{Mass spectra of bosons and fermions in Planck units
          in Model I (left) and Model II (right) on a logarithmic
          scale, in comparison with the compactification scale. The fermions are depicted right of the bosons.}
	\label{massplot}
\end{figure}

It is instructive to compare the mass spectra of the two models. From Eq.\ \eqref{chargedmass} one reads off that in both cases $m_+ \propto L^{-1}$. Hence, even the lightest charged scalar does not belong to the low-energy effective Lagrangian for fields with masses $m \ll L^{-1}$. Nevertheless, we included this scalar in the above discussion to check the stability of the vacuum, that is $m_+^2 > 0$. The vector boson, moduli and axion masses scale as $L^{-3/2}$ and become parametrically smaller than the size of the extra dimension for large $L$.

\section{Super-Higgs mechanism and fermion masses}
\label{sec:Fermions}
 
The first step in calculating the fermion masses is to disentangle
their mixing with the gravitino, i.e.\ to identify the
Goldstino. Since supersymmetry is broken by $F$- and $D$-terms, the
Goldstino is a mixture of the gaugino and modulini.
Charged fermions do not contribute as long as their scalar partners
have vanishing vacuum expectation values.
We can therefore restrict our discussion to the gaugino and modulini, and
we also assume a ground state with vanishing cosmological constant. 

The bilinear fermionic Lagrangian involves kinetic, mass, and mixing terms for the
gravitino, gaugino and modulini fields.
All these terms are determined by the K\"ahler potential, the
superpotential, the Killing vectors and the gauge kinetic function of
the model. The terms involving the gravitino $\psi_{\mu}$ are \cite{Wess:1992cp}
\begin{align}\label{Lgrav1}
\mathcal{L}_G =
\epsilon^{\mu\nu\rho\sigma}\gravb_\mu \sigb_\nu\partial_\rho\psi_\sigma
- m_{3/2} \psi_\mu\sigma^{\mu\nu}\psi_\nu - \overline{m}_{3/2} \gravb_\mu \sigb^{\mu\nu} \gravb_\nu
+ \psi_\mu\sigma^\mu \overline{\chi} + \chi\sigma^\mu \gravb_\mu\,, 
\end{align}
where $\chi$ is a linear combination of the fermions in the theory, and the gravitino mass is given by\footnote{In the following,
  bosonic terms are understood as expectation values in the ground state.}
\begin{align}
\overline{m}_{3/2} =  e^{K/2} W \,.
\end{align}
In the model under consideration this yields
\begin{align}
\overline{m}_{3/2} = \frac{1}{\sqrt{s t \tau_2}} \left(W_0 + W_1 \, e^{-a Z} + W_2 \, e^{- \tilde{a} U} \right) \,.
\label{gravmass}
\end{align}
The $D$-term affects the value of the gravitino mass via the expectation
values of the moduli fields determined in a vacuum with vanishing cosmological constant.

The fermion $\chi$ in Eq.~\eqref{Lgrav1} is the Goldstino. It is given by a linear combination of the gaugino $\lambda$ and modulini $\psi^i$, determined by $F$- and $D$-terms,
\begin{align}\label{chi}
\chi = -\frac{g}{2} D\,  \lambda - \frac{i}{\sqrt{2}}e^{K/2}D_i W \, \psi^i\,.
\end{align}
It is well known that the mixing terms in Eq.~\eqref{Lgrav1} can be
removed via a local field redefinition of the gravitino \cite{Wess:1992cp},
\begin{align}\label{shift}
\psi_\mu \rightarrow \psi_\mu -\frac{\sqrt{2}}{\sqrt{3} m_{3/2}}\partial_\mu\eta 
+ \frac{i}{\sqrt{6}}\overline{\eta}\,{\overline{\sigma}}_\mu\,, \quad \eta =
\frac{i\sqrt{2}}{\sqrt{3} \, \overline{m}_{3/2}}\chi\,.
\end{align}
With this shift, a straightforward calculation yields for the Lagrangian, \eqref{Lgrav1} 
\begin{align}\label{Lgrav2}
\mathcal{L}_G =
&\ \epsilon^{\mu\nu\rho\sigma} \gravb_\mu \sigb_\nu \partial_\rho \psi_\sigma
- m_{3/2} \psi_\mu \sigma^{\mu\nu} \psi_\nu - \overline{m}_{3/2} \gravb_\mu \sigb^{\mu\nu} \gravb_\nu \nonumber\\
& + i \overline{\eta} \, \sigb^\mu\partial_\mu\eta +
\overline{m}_{3/2}\eta\eta + m_{3/2}\overline{\eta}\overline{\eta}\,.
\end{align}
Note that the kinetic term of $\eta$ has the sign of a ghost. The
kinetic term and the mass term of $\eta$ lead to modifications of the kinetic terms and the mass
matrix of gaugino and modulini whereas the mass term for the gravitino
remains unchanged.

Eq.~\eqref{Lgrav2} represents the gravitino Lagrangian in unitary
gauge. Hence, the dependence on the Goldstino should completely
disappear in the full Lagrangian,
which we explicitly demonstrate in the following.
The kinetic terms of gaugino and modulini read
\begin{align}\label{Gmix}
\mathcal{L}_K = -i h \, \overline{\lambda} \sigb^\mu \partial_\mu \lambda -i K_{i \jb} \, \gravb^{\jb} \sigb^\mu \partial_\mu \psi^i\,.
\end{align}
Since the K\"ahler metric $K_{i \jb}$ is hermitian, it can be
diagonalized by a unitary transformation. Moreover, the real
eigenvalues are all positive for the kinetic terms to be well-defined. Hence, one can rescale the eigenvalues to one by conjugation
with a real diagonal matrix. The combined transformation corresponds
to a vielbein for the K\"ahler metric,
\begin{align}
g^i_{\; k} \, K_{i \jb} \, g^{\jb}_{\; \bar{l}} = \delta_{k \bar{l}} \,.
\end{align}
It is convenient to redefine gaugino and modulini,
\begin{align}
\lambda \rightarrow h^{-1/2} \lambda\,,\quad \psi^i \rightarrow
g^i_{\; k} \psi^k\,,
\label{rescaleferm}
\end{align}
such that their kinetic terms are canonical,
\begin{align}
\mathcal{L}_K = -i \overline{\lambda} \, \sigb^{\mu} \partial_{\mu} \lambda - i \delta_{i \jb} \, \gravb^{\jb} \, \sigb^{\mu} \partial_{\mu} \psi^i \equiv -i \delta_{a \bar{b}} \, \overline{\chi}^{\bar{b}} \, \sigb^{\mu} \partial_{\mu} \chi^{a} \,;
\end{align}
the new index $a$ labels gaugino ($\chi^0 = \lambda$) and modulini
($\chi^i = \psi^i$). One can now perform a unitary transformation
which rotates the canonically normalized fermions $(\lambda,\psi^i)$ into $\eta$ and three orthogonal Weyl fermions $\chi^i_\perp$
\begin{align}
\chi^a = U^{a}_{\,i} \, \chi^i_\perp + U^a_{\, \eta} \, \eta \,.
\label{unitrans}
\end{align}
From Eqs.~\eqref{chi} and \eqref{shift} one obtains for the matrix 
elements $(U^*)^{\eta}_{\; a}$ of the inverse transformation, 
\begin{align}\label{Umeta}
(U^{-1})^{\eta}_{\; 0} = (U^*)^{\eta}_{\; 0} = -\frac{i g D}{\sqrt{6 h} \, \overline{m}_{3/2}} \,,\quad
(U^{-1})^{\eta}_{\; i} = (U^*)^{\eta}_{\; i} = \frac{e^{K/2}D_k W}{\sqrt{3} \overline{m}_{3/2}} g^k_{\; i} \,.
\end{align}
One easily verifies the unitarity condition for the $\eta$-component,
\begin{align}
U^a_{\; \eta} (U^{-1})^{\eta}_{\; a} = \frac{1}{3 e^K |W|^2} \left( \frac{g^2}{2 h} D^2 + e^K K^{i \jb} D_i W D_{\jb} \Wb \right) \,,
\end{align}
where we used $K^{i \jb} = g^i_{\; k} \delta^{k \bar{l}} g^{\jb}_{\;\bar{l}}$. 
Clearly, in Minkowski space one has $U^a_{\; \eta} (U^{-1})^{\eta}_{\;a} = 1$.
Since all fermions are canonically normalized the kinetic terms for the orthogonal Weyl fermions $\chi^i_\perp$ and $\eta$ are given by
\begin{align}\label{Lperp}
\mathcal{L}_K =
-i\overline{\chi}^i_\perp \sigb^\mu \partial_\mu \chi^i_\perp 
-i\overline{\eta} \, \sigb^\mu \partial_\mu \eta \,.
\end{align}
Combined with Eq.~\eqref{Lgrav2} the kinetic terms for $\eta$ cancel, as expected in the unitary gauge.

Furthermore, we expect the mass eigenvalue of the Goldstino and its mixing
terms with the other fermions to vanish. In order to show this, we
express the  fermion bilinears in terms of the canonically normalized
gaugino and modulini,
\begin{align}
\mathcal{L}_M = -\tfrac{1}{2}M_{00} \, \lambda\lambda - \tfrac{1}{2} M_{km} \, \psi^k \psi^m - M_{0 k} \, \lambda \psi^k + \text{h.c.} = - \tfrac{1}{2} M_{ab} \chi^a \chi^b + \text{h.c.}
\end{align}
The mass matrix elements are given in App.~\ref{Appfermions}. Inserting \eqref{unitrans}, one obtains for the
mass term and mixing of the Goldstino field the expressions
\begin{equation}
\begin{split}
\tfrac{1}{2} M_{ab} \chi^a \chi^b &= \tfrac{1}{2} M_{ab} (U^a_{\;i} \chi^i_{\perp} +U^a_{\eta} \eta)(U^b_{\;j} \chi^j_{\perp} +U^b_{\eta} \eta) \\
&= \tfrac{1}{2} U^a_{\eta} M_{ab} U^b_{\eta} \, \eta \eta  + U^a_{\; i} M_{ab} U^b_{\eta} \, \eta \chi^i_{\perp} + \tfrac{1}{2} U^a_{\; i} M_{ab} U^b_{\; j} \, \chi^i_{\perp} \chi^j_{\perp} \,.
\label{fermionmassrotated}
\end{split}
\end{equation}
Using the explicit form of the fermion mass matrix and Eqs.~\eqref{Umeta}, a
straightforward calculation yields (see App.~\ref{Appfermions})
\begin{align}
M_{0b} U^b_{\eta} \propto 
\frac{D}{3 \overline{m}_{3/2}}\left(e^K\left(K^{i\jb} D_iW D_{\jb}\Wb -
    3|W|^2\right) 
+ \frac{g^2}{2h} D^2\right) + i\overline{X}^{\jb} \partial_{\jb}\Wb e^{K/2} \,,
\end{align}
which indeed vanishes in Minkowski space for a gauge invariant
superpotential. Analogously, one finds
\begin{align}
M_{k b} U^b_{\eta} \propto 
\partial_n V
-\frac{2}{3 W} D_n W \left( e^K \left( K^{\jb m} D_m W D_{\jb} \Wb -3 |W|^2\right)+ \frac{g^2}{2h} D^2 \right)  \,,
\end{align}
which also vanishes in Minkowski space for an extremum of the scalar
potential. Hence, the Goldstino indeed decouples from the mass matrix.

In summary, we obtain from Eqs.~\eqref{Lgrav2}, \eqref{Lperp} and
\eqref{fermionmassrotated},
\begin{align}\label{Lgravphys}
\mathcal{L}_G =
&\ \epsilon^{\mu\nu\rho\sigma} \gravb_\mu \sigb_\nu \partial_\rho \psi_\sigma
-i\overline{\chi}^i_\perp \sigb^\mu \partial_\mu \chi^i_\perp \nonumber\\
&- m_{3/2} \psi_\mu \sigma^{\mu\nu} \psi_\nu - \overline{m}_{3/2} \gravb_\mu \sigb^{\mu\nu} \gravb_\nu 
- \tfrac{1}{2} M_{a b}^{\perp}\, \chi^i_{\perp} \chi^j_{\perp} \,,
\end{align}
where 
\begin{align}
 M_{a b}^{\perp} = U^a_{\; i} M_{a b} U^b_{\; j} \,
\end{align}
is the mass matrix of the fermions orthogonal to the Goldstino. Again, the cancellation of $F$- and $D$-terms for small cosmological constant allows to derive a general scaling behavior for the fermion masses. With Eqs.~\eqref{Dtermscaling} and \eqref{superpotscaling} we obtain
\begin{align}
m_{3/2} \propto L^{-3/2} \,, \quad M_{\text{modulini}} \propto L^{-3/2} \,.
\label{fermassscaling}
\end{align}

\subsubsection*{Gravitino and fermion masses: two examples}

Using the expression \eqref{gravmass} for the gravitino mass one
obtains for the two models defined by Eqs.~\eqref{vacua} and the
vacuum values of the moduli fields $(s,t)_{\text{I,II}} = (\kappa, \kappa^{-1})_{\text{I,II}}$,
\begin{align}
\left(m_{3/2}\right)_{\text{I}} \approx 1.8 \times 10^{-3} \,, \quad \left( m_{3/2} \right)_{\text{II}} \approx 5.2 \times 10^{-9} \,.
\end{align}
The modulini masses are evaluated numerically from Eq.~\eqref{condition1}.
As expected, one finds one eigenvector with
zero mass, the Goldstino.
The three remaining fermion fields are massive with mass eigenvalues of order the gravitino mass
\begin{align}
M_{\text{modulini}} = \mathcal{O}(m_{3/2}) \,.
\end{align}
Again, the explicit numerical values and scaling behaviors are summarized in App.~\ref{appnumerical}. An overview over the mass spectrum in both vacua is provided in Fig.~\ref{massplot}. Interestingly, in both of the vacua one modulini field remains lighter than the gravitino whereas the other two are slightly heavier.

\section{Summary and Outlook}
\label{sec:Conclusion} 
We have analyzed a 6d $\mathcal{N}=1$ supergravity model compactified
to 4d on an orbifold. Using bulk flux, a nonperturbative
superpotential, and the Green-Schwarz term for anomaly cancellation
we obtained 4d de Sitter or Minkowski vacua where all moduli are
stabilized. This allows for an explicit computation of the masses of
all particles in the effective low-energy theory.

In the model under discussion, supersymmetry is broken by both $F$-
and $D$-terms. By analyzing the bosonic 6d effective action, we extracted the $D$-term potential resulting from the FI parameter of
the anomalous $U(1)$, which 
receives contributions from the Green-Schwarz term and from the bulk flux. From the Green-Schwarz term we also obtained an important correction to the gauge
kinetic function.
The $F$-term potential results from our choice of the superpotential
which is of the KKLT-type. Knowing the complete scalar potential we
then calculated the boson masses which depend on the superpotential parameters and the flux.

For the discussion of the fermion masses we have studied the super-Higgs
mechanism in the presence of $F$- and $D$-term breaking. Via a
rotation in field space we extracted the Goldstino which is eaten
by the gravitino in unitary gauge. The Goldstino indeed completely 
drops out of the Lagrangian, as we explicitly verified using the
extremum conditions of the scalar potential for Minkowski space and the gauge invariance of
the superpotential.

In order to find vacua of our effective theory that are de Sitter or
Minkowski and are within a reasonable parameter range for the moduli,
we inverted the problem: Choosing a gauge coupling and the size of the
extra dimensions, and starting from a point in moduli space 
we derived equations for the superpotential parameters for which the
scalar potential is minimized. Having obtained the parameters of our
effective theory that way, we inserted the parameters back into the
scalar potential which we then minimized. As a cross-check, we found the minimum exactly at the point in moduli space which we used to obtain the parameters in the inverted problem.

Finally, we discussed two example models with different parameters
and evaluated numerically the masses of all particles in the
model. In the first example, the extra dimensions are of order the GUT
scale, $L^{-1} \sim 10^{-2}\Mp$, and moduli, axion and gauge boson
masses are also $\mathcal{O}(10^{-2} \Mp)$, slightly larger than
gravitino and modulini masses. In the second example, the size of the
extra dimensions corresponds to an intermediate scale, $L^{-1} \sim
10^{-6}\Mp$ and all masses scale as $m^2 \sim L^{-3}$. The only exception is the charged scalar mass which is of the order of the compactification scale. The size of the
extra dimensions is controlled by the gauge coupling, with $g^2 L =
\mathcal{O}(10)$. This dependence on the gauge coupling can be used to
construct models whose size of the extra dimensions interpolates
between the GUT scale and the TeV scale. 

The constructed family of de Sitter vacua can easily be combined with
higher-dimensional GUT models, and they also offer an interesting
playground to study the interplay of moduli stabilization and
inflation. Since the considered 6d flux compactifications contain all
ingredients familiar from string models, i.e. compact dimensions with
flux, the Green-Schwarz mechanism and a nonperturbative
superpotential, it will be very interesting to see whether they can in
fact be realized within a string theory construction.

\section*{Acknowledgments}
We thank Emilian Dudas, Zygmunt Lalak, Jan Louis, Hans-Peter Nilles,
and Alexander Westphal for valuable discussions. This work was supported by the German Science Foundation (DFG) within the Collaborative Research Center (SFB) 676 ``Particles,
Strings and the Early Universe''. M.D. also acknowledges support from the Studienstiftung des deutschen Volkes.

\begin{appendix}

\section{Parameters for de Sitter and Minkowski vacua}
\label{parameters}

For the evaluation of the superpotential parameters it is convenient to define new linear combinations of the derivative operators
\begin{align}
\partial_+ = s \partial_S + t \partial_T \,, \quad \partial_- = s \partial_S - t \partial_T \,, \quad \partial_0 = \tau_2 \partial_U \,.
\label{eqsysmod}
\end{align}
In terms of this derivative operators the constraints \eqref{constraints} can be rewritten as (note that $s,t,\tau_2 > 0$)
\begin{align}
\partial_+ V = 0 \,, \quad \partial_- V = 0 \,, \quad \partial_0 V = 0 \,, \quad V = \epsilon \,.
\end{align}
Using the specific form of the nonperturbative superpotential \eqref{superpot} the parameters can be identified as:
\begin{subequations}
\begin{align}
A &= a^2 \, W_1^2 \, e^{-a z} \,, \label{eqA} \\
\tilde{A} &= \tilde{a}^2 \, W_2^2 \, e^{-\tilde{a} \tau_2} \,, \label{eqAtilde} \\
B &= -2 a W_0 W_1 \, e^{-\frac{a}{2} z} \cos \left( \tfrac{a}{2} \tilde{c} \right) - 2 a W_1^2 \, e^{-a z} - 2 a W_1 W_2 \, e^{- \frac{a}{2} z - \frac{\tilde{a}}{2} \tau_2} \cos \left( \tfrac{a}{2} \tilde{c} - \tfrac{\tilde{a}}{2} \tau_1 \right) \,,\\
\tilde{B} &= -2 \tilde{a} W_0 W_2 \, e^{-\frac{\tilde{a}}{2} \tau_2} \!\cos \left( \tfrac{\tilde{a}}{2} \tau_1 \right) \!-\! 2  \tilde{a} W_2^2 \, e^{- \tilde{a} \tau_2} \!-\! 2 \tilde{a} W_1 W_2 \, e^{- \frac{a}{2} z - \frac{\tilde{a}}{2} \tau_2} \!\cos \left( \tfrac{a}{2} \tilde{c} - \tfrac{\tilde{a}}{2} \tau_1 \right) \,.
\end{align}
\end{subequations}
Under the assumption $W_0 < 0$ and $W_1, W_2 > 0$ the imaginary parts are generically stabilized at zero. Nevertheless, this has to be checked a posteriori for the specific superpotential parameters. In the following we will therefore set $\tilde{c} = 0$ and $\tau_1 = 0$ and crosscheck for consistency afterward. We further introduce a convenient new combination
\begin{align}
C = a \tilde{a} W_1 W_2 \, e^{-\frac{a}{2} z - \frac{\tilde{a}}{2} \tau_2} = \partial_U W \, \partial_{\overline{Z}} \Wb = \partial_Z W \, \partial_{\overline{U}} \Wb \,.
\end{align}
Using the relations
\begin{align}
\partial_+ A = -ast E A \,, \enspace \partial_+ B = stE \left(A - \tfrac{a}{2} B \right) \,, \enspace \partial_+ \tilde{A} = 0 \,, \enspace \partial_+ \tilde{B} = s t E C \,,
\end{align}
and analogous equations for $\partial_-$, where $E$ is substituted by $D$, as well as
\begin{align}
\partial_0 A = 0 \,, \enspace \partial_0 B = \tau_2 C \,, \enspace \partial_0 \tilde{A} = - \tilde{a} \tau_2 \tilde{A} \,, \enspace \partial_0 \tilde{B} = \tau_2 \left(\tilde{A} - \tfrac{\tilde{a}}{2} \tilde{B}\right) \,,
\end{align}
one finds
\begin{subequations}
\begin{align}
\partial_+ V &= -\tfrac{(st)^2}{2 \tau_2} a E A (D^2 + E^2) - \tfrac{st}{\tau_2} E^2 \left( A - \tfrac{a}{2} B \right) + \tfrac{1}{\tau_2} B E - \tfrac{2 \tau_2}{st} \tilde{A} + \tfrac{2}{st} \tilde{B} - E C \nonumber\\ &\quad\;\!-\tfrac{3g^2}{2 h} D^2 \,,\\
\partial_- V &= \tfrac{2st}{\tau_2} A E D - \tfrac{(st)^2}{2 \tau_2}a D A (D^2 + E^2) - \tfrac{st}{\tau_2} DE \left( A - \tfrac{a}{2} B\right) - \tfrac{1}{\tau_2} BD - DC \nonumber \\ 
&\quad\;\! - \tfrac{\tilde{h} g^2}{2 h^2} D^2 + \tfrac{g^2}{h} D E \,, \\
\partial_0 V &= -\tfrac{st}{2 \tau_2} A (D^2 + E^2) + \tfrac{1}{\tau_2} BE - CE - \tfrac{\tilde{a} \tau_2^2}{st} \tilde{A} + \tfrac{\tilde{a} \tau_2}{2 st} \tilde{B} \,.
\end{align}
\end{subequations}
Here, we have introduced the linear combination $\tilde{h} =\frac12\left(h_S s - h_T t\right)$.

One can then reverse the problem of finding a minimum of the scalar potential for fixed parameters by fixing the values of the moduli and trying to solve Eqs.~\eqref{eqsysmod} for the superpotential parameters instead. It is obvious that the superpotential \eqref{superpot} has five free parameters, but we only have four equations to determine them. Therefore, generically we have to fix one of the parameters in order to solve for the others\footnote{It is a natural choice to fix $\tilde{a}$ which corresponds to nonperturbative effects in $U$.}.

The solution for the superpotential parameters has to fulfill several consistency conditions. For instance, taking into account Eqs.~\eqref{eqA} and \eqref{eqAtilde}, $A$ and $\tilde{A}$ have to be positive. Interestingly, the combination of the equations $D \partial_+ V - E \partial_- V = 0$ with the constraint that $V = \epsilon$, fixes $A$ unambiguously
\begin{align}
A = \frac{g^2 \tau_2}{2 st h^2 (\rho^2 - 1)} \left( h_T t \rho + h (2 - \rho + \rho^2) + \frac{4 h^2 \epsilon}{g^2 E^2}\right) \,,
\end{align}
where we have introduced the new variable $\rho = D/E$. Unfortunately, the rest of the equations can not be simplified in a similar way and have to be evaluated numerically.

\section{Fermion masses}
\label{Appfermions}

In the following we derive the mass matrix for canonically normalized
fermions and show explicitly that the Goldstino decouples in the unitary gauge.

The standard fermion mass terms for gaugino and modulini are given in full generality in \cite{Wess:1992cp}. For our case they read
\begin{equation}
\begin{split}
\mathcal{L}_{M} =& \sqrt{2} g K_{i \jb} \overline{X}^{\jb} \psi^i \lambda - \frac{ig}{\sqrt{2}} \frac{\partial_i h}{h} D \psi^i \lambda \\
&- \tfrac{1}{2} e^{K/2} (\mathcal{D}_i D_j W) \psi^i \psi^j + \tfrac{1}{2} e^{K/2} K^{i \jb} (D_{\jb} \Wb) (\partial_i h) \lambda \lambda + \text{h.c.} \,,
\end{split}
\end{equation}
where $\mathcal{D}_i D_j W = W_{i j} + K_{i j} W + K_{i} D_j W + K_j
D_i W - K_i K_j W - \Gamma^{k}_{ij} D_k W$ (with $W_{ij}
= \partial_i \partial_j W$ etc.). After the rescaling to canonical
kinetic terms \eqref{rescaleferm} and transforming to the unitary
gauge \eqref{shift}, the fermion bilinears are given by
\begin{equation}
\begin{split}
\mathcal{L}_{M} =& \frac{\partial_i h}{2 h} e^{K/2} K^{i \jb} (D_{\jb} \Wb) \lambda \lambda - \frac{1}{2} e^{K/2} (\mathcal{D}_i D_j W) g^i_{\; k} g^j_{\; m} \psi^k \psi^m \\
& + \frac{\sqrt{2} g}{\sqrt{h}} K_{i \jb} \overline{X}^{\jb} g^i_{\; k} \psi^k \lambda - \frac{i g}{\sqrt{2}} \frac{\partial_i h}{h^{3/2}} D g^i_{\; k} \psi^k \lambda  + \overline{m}_{3/2} \eta \eta + \text{h.c.}\\
=& - \tfrac{1}{2} M_{0 0} \, \lambda \lambda- \tfrac{1}{2} M_{k m} \, \psi^k \psi^m - M_{0 k} \, \psi^k \lambda + \text{h.c.} \\\equiv& - \tfrac{1}{2} M_{a b} \, \chi^a \chi^b + \text{h.c.} \,.
\end{split}
\end{equation}
Using $K_{i\jb} \Xb^{\jb} = i \partial_i D$ \cite{Wess:1992cp}, the mass matrix elements read
\begin{equation}
\begin{split}
M_{00} &= - \frac{\partial_i h}{h} e^{K/2} K^{i \jb} (D_{\jb} \Wb) + \frac{g^2}{3 h \overline{m}_{3/2}} D^2  \\
&= - \frac{1}{h} e^{- K/2} \left( e^K K^{i \jb} \, \partial_i h \, D_{\jb} \Wb - \frac{1}{3 W} g^2 D^2 \right) \,, \\
M_{0k} &= - \frac{\sqrt{2} g}{\sqrt{h}} K_{i \jb} \overline{X}^{\jb} g^i_{\; k} + \frac{i g}{\sqrt{2h}} \frac{\partial_i h}{h} D g^i_{\;k} + \frac{\sqrt{2} i g}{3 \sqrt{h} W} D_i W D g^i_{\; k} \\
&= - i \frac{\sqrt{2} g}{\sqrt{h}} g^i_{\; k} \left( \partial_i D -  \frac{\partial_i h}{2 h} D - \frac{1}{3 W} D_i W D  \right) \,, \\
M_{km} &= e^{K/2} (\mathcal{D}_i D_j W) g^i_{\; k} g^j_{\; m} - \frac{2}{3 W^2} \overline{m}_{3/2} D_i W D_j W g^i_{\;k} g^j_{\;m} \\
&=  e^{K/2} g^i_{\;k} g^j_{\;m} \left( \mathcal{D}_i D_j W - \frac{2}{3 W} D_i W D_j W \right)\,. \label{condition1}
\end{split}
\end{equation}
In order to show the decoupling of the Goldstino we now perform the unitary transformation to the $(\eta,
\chi^i_{\perp})$ basis, for which we have to show $M_{ab}U^b_{\eta} =
0$. Using the above expressions for $M_{ab}$ and Eq.~\eqref{Umeta} for
$U^b_{\eta}$ one obtains
\begin{equation}
\begin{split}
M_{0b} U^b_{\eta} &= \frac{1}{\sqrt{3}m_{3/2}}\left\{
M_{00} \frac{i g }{\sqrt{2h}} D + M_{0k} \delta^{k \bar{l}} g^{\jb}_{\;
    \bar{l}} D_{\jb} \Wb e^{K/2}\right\} \\
&=\frac{i\sqrt{2}}{\sqrt{3}m_{3/2}}\frac{g}{\sqrt{h}}\left\{\frac{D}{3\overline{m}_{3/2}}
\left(e^K K^{i\jb} D_iW D_{\jb}\Wb + \frac{g^2}{2h} D^2\right)
-\partial_i D K^{i\jb} D_{\jb}\Wb e^{K/2}\right\}\\
&=\frac{i\sqrt{2}}{\sqrt{3}m_{3/2}}\frac{g}{\sqrt{h}}\left\{\frac{D}{3\overline{m}_{3/2}}
\left(e^K K^{i\jb} D_iW D_{\jb}\Wb + \frac{g^2}{2h} D^2\right) \right.\\
&\hspace*{4cm}+ i\overline{X}^{\jb} \partial_{\jb}\Wb e^{K/2} - D\Wb e^{K/2}\bigg\}\,,
\end{split}
\end{equation} 
where we used $\partial_i D = -i K_{i\jb}\overline{X}^{\jb}$. Clearly,
the gauge invariance of the superpotential, i.e $\overline{X}^{\jb} \partial_{\jb} \Wb = 0$,
implies that in Minkowski space
\begin{align}
M_{0b}U^b_{\eta} = 0\,.
\end{align}
Similarly, one finds for the  second condition
\begin{equation}
\begin{split}
M_{k b} U^{b}_{\; \eta} =& \frac{1}{\sqrt{3}m_{3/2}}\left\{
M_{0k} \frac{i g D}{\sqrt{2h}} + M_{i k} \delta^{\bar{m} i}
g^{\jb}_{\; \bar{m}} D_{\jb} \Wb e^{K/2}\right\} \\
=&  \frac{1}{\sqrt{3}m_{3/2}} g^n_{\; k} \left\{ e^K K^{\jb m} \mathcal{D}_m D_n W D_{\jb} \Wb + \frac{g^2}{h} D \partial_n D - \frac{g^2}{2 h^2} \partial_n H D^2 \right. \\
& \left.  \hspace*{2cm} - \frac{2}{3 W} D_n W \left( e^K K^{\jb m} D_m W D_{\jb} \Wb + \frac{g^2}{2h} D^2 \right) \right\} \\
=&  \frac{1}{\sqrt{3}m_{3/2}} g^n_{\; k} \left\{ \partial_n V_F
  + \partial_n \left(\frac{g^2}{2 h} D^2 \right) + 2 e^K D_n W \Wb 
\right. \\
& \left.  \hspace*{2cm} -\frac{2}{3 W} D_n W \left( e^K K^{\jb m} D_m W D_{\jb} \Wb + \frac{g^2}{2h} D^2 \right) \right\} \,,
\label{condition2}
\end{split}
\end{equation}
where we have used the identity 
\begin{align}
e^K K^{m\jb} \mathcal{D}_m D_n W D_{\jb} \Wb = \partial_n V_F  + 2 e^K D_n W \Wb \,.
\end{align}
From Eq.~\eqref{condition2} one reads off that 
\begin{align}
M_{kb}U^b_{\eta} = 0 
\end{align}
for an extremum of the potential and vanishing vacuum energy density.

As expected, for a gauge invariant superpotential the Goldstino completely decouples from the mass matrix
at an extremum of the potential with vanishing cosmological
constant. In the case of a small cosmological constant $\epsilon$ the
above equations are modified by terms $\mathcal{O}(\epsilon)$.

\section{Numerical evaluation of masses}
\label{appnumerical}

We summarize the superpotential parameters and masses of the bosons and fermions in the two vacua \eqref{vacua} with different sizes of the extra dimensions and demonstrate the predicted scaling behavior. The vacuum expectation values for the moduli in the vacuum are
\begin{align}
(s,t,u)_{\text{I}} = \left(\tfrac{3}{5},\tfrac{5}{3},1\right) \,, \quad (s,t,u)_{\text{II}} = \left(\tfrac{6}{5},\tfrac{5}{6},1\right) \,.
\end{align}
The other input parameters for the two models are
\begin{align}
g_{\text{I}} = 0.2 \,, \enspace L_{\text{I}} = 200 \,, \quad g_{\text{II}} = 4 \times 10^{-3} \,, \enspace L_{\text{II}} = 10^{6} \,,
\end{align}
corresponding to a scale parameter of $L_{\text{I}}/L_{\text{II}} = 5 \times 10^3$. In order to obtain unique solutions for the superpotential we have to fix one of its parameters, see App.~\ref{parameters}. Therefore, we choose $(\tilde{a})_{\text{I}} = 2$ and $(\tilde{a})_{\text{II}} = 3$. The numerical values for the superpotential parameters and their scaling behavior are summarized in Tab.~\ref{superparatab}.

With the parameters of the superpotential the moduli, axion, and modulini masses can also be calculated numerically (after canonical normalization of the kinetic term). The mass eigenvalues are summarized in Tab.~\ref{massestab}. The scaling with respect to the size of the extra dimensions matches nicely the expressions in \eqref{bosmassscaling} and \eqref{fermassscaling}. Both mass hierarchies are depicted in a logarithmic scale in Fig.~\ref{massplot}.

\renewcommand{\arraystretch}{1.3}
\begin{table}[ht]
\centering
\begin{tabular}{|c||c|c|c|}
\hline
 & Model I & Model II & Scaling\\ 
\hline \hline
$V_2^{-1/2}$ & $7.1 \times 10^{-3}$ & $1.4 \times 10^{-6} $ & $5.0 \times 10^3$ \\
\hline
$W_0$ & $-3.0 \times 10^{-3}$ & $-7.4 \times 10^{-9}$ & $5.5 \times 10^3$ \\
\hline
$W_1$ & $1.2 \times 10^{-2}$ & $3.3 \times 10^{-8}$ & $5.0 \times 10^3$ \\
\hline
$W_2$ & $2.8 \times 10^{-3}$ & $8.5 \times 10^{-9}$ & $4.8 \times 10^3$ \\
\hline 
$a$ & $4.5 \times 10^{2}$ & $2.3 \times 10^{6}$ & $5.0 \times 10^3$ \\
\hline 
$a z$ & $9.4$ & $9.4$ & \\
\hline
$\tilde{a}$ & $2$ & $3$ & \\
\hline
\end{tabular}
\caption{Numerical values of the superpotential parameters in Planck units for the two models \eqref{vacua}. The scaling parameter is evaluated according to the behavior predicted in Eq.~\eqref{superpotscaling}.}
\label{superparatab}
\end{table}
\begin{table}[ht]
\centering
\begin{tabular}{|c||c|c|c|}
\hline
 & Model I & Model II & Scaling\\ 
\hline \hline
$V_2^{-1/2}$ & $7.1 \times 10^{-3}$ & $1.4 \times 10^{-6} $ & $5.0 \times 10^3$ \\
\hline
$m_{3/2}$ & $1.8 \times 10^{-3}$ & $5.2 \times 10^{-9}$ & $5.0 \times 10^3$ \\
\hline
$m_+$ & $4.2 \times 10^{-2}$ & $8.3 \times 10^{-6}$ & $5.0 \times 10^3$ \\
\hline
$m_A$ & $1.1 \times 10^{-2}$ & $3.0 \times 10^{-8}$ & $5.0 \times 10^3$ \\
\hline 
$m_{\text{modulus},1}$ & $1.2 \times 10^{-2}$ & $3.4 \times 10^{-8}$ & $5.0 \times 10^3$ \\
\hline 
$m_{\text{modulus},2}$ & $4.1 \times 10^{-3}$ & $1.5 \times 10^{-8}$ & $4.1 \times 10^3$ \\
\hline 
$m_{\text{modulus},3}$ & $3.2 \times 10^{-3}$ & $1.1 \times 10^{-8}$ & $4.3 \times 10^3$ \\
\hline 
$m_{\text{axion},1}$ & $9.4 \times 10^{-3}$ & $2.7 \times 10^{-8}$ & $4.9 \times 10^3$ \\
\hline
$m_{\text{axion},2}$ & $4.0 \times 10^{-3}$ & $1.6 \times 10^{-8}$ & $3.8 \times 10^3$ \\
\hline
$M_{\text{modulini},1}$ & $4.8 \times 10^{-3}$ & $1.8 \times 10^{-8}$ & $4.1 \times 10^3$ \\
\hline
$M_{\text{modulini},2}$ & $2.6 \times 10^{-3}$ & $8.3 \times 10^{-9}$ & $4.7 \times 10^3$ \\
\hline
$M_{\text{modulini},3}$ & $7.4 \times 10^{-4}$ & $2.0 \times 10^{-9}$ & $5.0 \times 10^3$ \\
\hline
\end{tabular}
\caption{Numerical values of masses in Planck units for gravitino, charged scalar, vector boson, moduli, axions and modulini. The scaling parameter is evaluated according to the behavior predicted in Eqs.~\eqref{bosmassscaling} and \eqref{fermassscaling}.}
\label{massestab}
\end{table}
\renewcommand{\arraystretch}{1.0}

\end{appendix}

\newpage


\providecommand{\href}[2]{#2}

\end{document}